\documentstyle[12pt]{article}
\begin{document}
\baselineskip 16pt plus 2pt minus 2pt
\newcommand{\beq}{\begin{equation}}
\newcommand{\eeq}{\end{equation}}
\newcommand{\beqa}{\begin{eqnarray}}
\newcommand{\eeqa}{\end{eqnarray}}
\newcommand{\dfrac}{\displaystyle \frac}
\renewcommand{\thefootnote}{\#\arabic{footnote}}
\newcommand{\ve}{\varepsilon}
\newcommand{\krig}[1]{\stackrel{\circ}{#1}}
\newcommand{\barr}[1]{\not\mathrel #1}
\begin{titlepage}

\begin{center}

\vspace{2.0cm}

{\large  \bf { BARYON CHIRAL PERTURBATION THEORY A.D. 1994}}

\vspace{1.2cm}

{\large  Ulf-G. Mei\ss ner$^{\dag,\star}$}

\vspace{0.7cm}

$^{\dag}$Universit\"at Bonn, Institut f\"ur Theoretische Kernphysik, Nussallee
14-16,\\ D--53115 Bonn, Germany

\vspace{0.5cm}

$^\star$email: meissner@pythia.itkp.uni-bonn.de

\end{center}

\vspace{2.5cm}

\begin{center}

{\bf ABSTRACT}

\end{center}

\vspace{0.1cm}

\noindent
In these lectures, the status of baryon chiral perturbation theory is
reviewed. Particular emphasis is put on the two--flavor sector and the
physics related to electromagnetic probes. I discuss in some detail
the structure of the effective Lagrangian at next--to--leading order,
the meaning of low--energy theorems in Compton scattering and pion
photoproduction and confront the chiral predictions with the existing data.
Some remaining problems and challenges are outlined.

\vspace{3.5cm}

\begin{center}

Lectures given at the Indian Summer School on Electron Scattering
off Nucleons and Nuclei, Prague, Czech Republic, September 1994

\end{center}

\vspace{2cm}

\vfill

\noindent TK 94 17 \hfill November 1994

\end{titlepage}
\section{Introduction}
\label{sec:intro}
Over the last few years, a tremendous amount of very precise data
probing the structure of the nucleon
at low energies has become available. Just to
name a few, these are the measurements of neutral pion photo-- and
electroproduction in the threshold region \cite{sac} \cite{beck}
\cite{pat} or the determinations of the nucleon electromagnetic
polarizabilities at
Illinois, Saskatoon, Oak Ridge and Mainz \cite{feder} \cite{schmi}
\cite{sal} \cite{zieger}. While the interpretation of some of these
data has been controversial, it is unchallenged that they encode
information about the nucleon in the non--perturbative regime, i.e.
at typical momentum scales were straightforward perturbation theory
in the running strong coupling constant $\alpha_S (Q^2)$ is no longer
applicable. At
present, there exist essentially two approaches to unravel the physics
behind the wealth of empirical information. On one hand, one uses
models which stress certain aspects of the strong interactions
(but tend to neglect or forget about others) like e.g. the constituent
quark model presented here by Buchmann \cite{alfons}. The other
possibilitiy is to make use of the symmetries of the Standard Model (SM)
and formulate an effective field theory (EFT) to systematically  explore
the strictures of these symmetries. In the case of QCD and in the
sector of the three light quarks u, d and s we know that there exist
an approximate chiral symmetry which is spontaneously broken with the
apperance of nine Goldstone bosons, the pions, kaons and the eta. These
pseudoscalar mesons are the lightest strongly interacting particles
and their small but non--vanishing masses can be traced back to
the fact that the current masses $m_u$, $m_d$ and $m_s$ are small
compared to any typical hadronic scale, like e.g. the proton mass.
In the meson sector, the EFT is called chiral perturbation theory and
is well developed and applied  successfully to many reactions
\cite{GL1} \cite{GL2} (for a review, see \cite{UGM}). Notice that
Ecker has given a nice introduction to this topic at this school
last year \cite{ecker}
so I will try to avoid too much overlap with his lectures. Of course,
there is also the lattice formulation of QCD which over the years has
shown great progress but is not yet in the status to discuss in detail
what will be mostly the topic here, namely the baryon structure as
accurately probed with real or virtual photons. Clearly, the baryons
(and in particular the nucleons I will mostly focus on) are not related
directly to the spontaneous chiral symmetry breakdown. However, their
interactions with pions and among themselves are strongly constrained
by chiral symmetry. This is, of course, known since the sixties (see
e.g. the lectures by Coleman \cite{col} and references
therein). However, to go beyond the current algebra or tree level
calculations, one needs a systematic power counting scheme as it was
first worked out for the meson sector by Weinberg \cite{wein79}. As
shown by Gasser et al. \cite{GSS}, the straightforward generalization
to the baryon sector leads to problems related to the non--vanishing
mass of the baryons in the chiral limit, i.e. one has an extra large
mass scale in the problem. Stated differently, baryon four--momenta
are never small compared to the chiral symmetry breaking scale
$\Lambda_\chi$, $m_B / \Lambda_\chi \sim 1$. This can be overcome by a
clever choice of velocity--dependent fields \cite{JM} which allows to
transform the baryon mass term in a string of $1/m_B$ suppressed
interaction vertices. Then, a consistent power counting scheme emerges
where the expansion in small momenta and quark masses can be mapped
one--to--one on a (Goldstone boson) loop expansion.

These lectures will be organized as follows. After a short review of
the construction of the EFT in the presence of matter fields, I will
give a sample calculation (of the pion cloud contribution to the
nucleon mass). I elaborate in some detail on the structure of the
next--to--leading order terms of dimension two in the effective
Lagrangian. In particular, I show how certain operators with {\it
 fixed} coefficients arise when the baryons are treated in the extreme
non--relativistic limit. This is followed by a study of the
numerical values of the low--energy constants appearing at this order.
Then, I will discuss the status of the low--energy
theorems in pion photo-- and electroproduction and Compton scattering.
In section 7, chiral predictions are confronted with the data and the
problems and open questions are adressed in section 8.

\section{Chiral perturbation theory with baryons}
\label{sec:CHPT}
Chiral perturbation theory (CHPT) is the EFT of the SM at low
energies in the hadronic sector. Since as an EFT it contains all terms allowed
by the symmetries of the underlying theory \cite{wein79}, it should be viewed
as a direct consequence of the SM itself. The two main assumptions
underlying CHPT are that
\begin{enumerate}
\item[(i)] the masses of the light quarks u, d (and possibly s) can be
treated as perturbations (i.e., they are small compared to a typical
hadronic scale of 1 GeV) and  that
\item[(ii)] in the limit of zero quark masses, the chiral symmetry is
spontaneously broken  to its vectorial subgroup. The resulting Goldstone
bosons are the pseudoscalar mesons (pions, kaons and eta).
\end{enumerate}

\noindent CHPT is a systematic low--energy expansion around the
chiral limit \cite{wein79} \cite{GL1} \cite{GL2} \cite{leut}. It is a
well--defined quantum field theory although it has to be renormalized
order by order. Beyond leading order, one has to include loop diagrams to
 restore unitarity perturbatively. Furthermore, Green functions calculated in
CHPT at a given order contain certain parameters that are not constrained by
the symmetries, the so--called low--energy constants (LECs).
At each order in the chiral expansion, those LECs have to be
determined  from phenomenology (or can be estimated with some
model dependent assumptions).
For a review of the wide field of applications of CHPT, see, e.g., Ref.~\cite
{UGM}.

In the baryon sector, a
complication arises from the fact that the baryon mass $m_B$ does not vanish
in the chiral limit \cite{GSS}. Stated differently, only baryon three--momenta
can be small compared to the hadronic scale. To see this in more
detail, let us consider the two--flavor case with $m_u = m_d = \hat
m$ and collect the proton and the neutron in a bi-spinor
\beq
\Psi  = \left( \begin{array}{c}
                 p \\ n
\end{array} \right)  \quad ,
\label{psi}
\eeq
which transforms non--linearly under chiral transformations,
\beq
\Psi \to \Psi' = K(L,R,U(x)) \, \Psi \quad .
\eeq
Here, $K(L,R,U(x))$ is a non--linear function of the meson fields
collected in $U(x)$ and of $L,R \in$ SU(2) \cite{ccwz}
\cite{weinno}. It is defined via
\beq
u' = L u K^\dagger = K u R^\dagger
 \, , \, {\rm with } \, \, U = u^2 \, , \, u'^2 = U'^2 \quad . \eeq
The unimodular unitary matrix $U$ ($U^\dagger U = U U^\dagger =1$,
det$(U)$ = 1) transforms linearly under chiral transformations,
\beq U \to U' = L U R^\dagger \quad . \eeq
It is most convenient to parametrize $U$ as follows
\beq U = (\sigma + i \vec \tau \cdot \vec \pi \, ) / F \, , \quad
\sigma^2 + {\vec \pi \,}^2 = F^2 \, \, , \eeq
with $F$ the pion decay constant in the chiral limit, $F_\pi = F[1
+{\cal O}(\hat m)] =93$ MeV. It is now
straightforward to write down chiral covariant derivatives and
construct the lowest order effective Lagrangian
\beq {\cal L}_{\rm eff}   = {\cal L}_{\pi N}^{(1)} +
 {\cal L}_{\pi \pi}^{(2)} \eeq
\beq {\cal L}_{\pi N}^{(1)}  = \bar{\Psi}
 \left( i \gamma_\mu D^\mu - \krig{m} + \frac{1}{2}\krig{g}_A \gamma^\mu
   \gamma_5 u_\mu \right) \Psi \label{lpin} \eeq
\beq  {\cal L}_{\pi \pi}^{(2)} =
\frac{F^2}{4} {\rm Tr} \left[ \nabla_\mu U \nabla^\mu U^\dagger
\right] + \frac{F^2 M^2}{4} {\rm Tr} \left[ U + U^\dagger \right]
 \quad , \eeq
with $u_\mu = i u^\dagger \nabla_\mu U u^\dagger$.
Here, the superscript '$\circ$' denotes quantities in the chiral limit, i.e.
$Q = ~\stackrel{\circ}{Q} [ 1 + {\cal O}(\hat m)]$
(with the exception of $M$ which is
the leading term in the quark mass expansion of the pion mass and $F$),
$g_A$ is the axial--vector
coupling constant measured in neutron $\beta$--decay, $g_A =
1.26$ and $m$ denotes the nucleon mass.
For the three flavor case, one has of course two axial
couplings. The chiral dimension (power) of the respective terms is
denoted by the superscripts '(i)' (i=1,2). The pertinent covariant
derivatives are
\beq \nabla_\mu U = \partial_\mu U - i e A_\mu [Q,U] \eeq
\beq D_\mu \Psi = \partial_\mu \Psi + \frac{1}{2} \lbrace u^\dagger
\left( \partial_\mu -i e A_\mu Q \right)u + u
\left( \partial_\mu -i e A_\mu Q \right)u^\dagger \rbrace \Psi
= \partial_\mu \Psi + \Gamma_\mu \Psi
\label{conn} \eeq
with $Q= {\rm diag}(1,0)$ the (nucleon) charge matrix and I only consider
external vector fields, i.e. the photon $A_\mu$. $D_\mu$ transforms
homogeneously under chiral transformation, $D_\mu' = K D_\mu
K^\dagger$, and  $\Gamma_\mu$ is the connection \cite{ccwz}.
 To show the strength of the effective Lagrangian approach,
let me quickly derive the so--called Goldberger--Treiman relation
(GTR) \cite{GTR}. For that, I set $A_\mu = 0$ and expand the
pion--nucleon Lagrangian to order $\vec \pi$,
\beq {\cal L}_{\pi N}^{(1)}  = \bar{\Psi}
 \left( i \gamma_\mu \partial^\mu - \krig{m} \right) \Psi
 - \frac{\stackrel{\circ}{g_A}}{F} \bar \Psi \gamma^\mu
   \gamma_5 \frac{\vec \tau}{2} \Psi \cdot \partial_\mu \vec \pi
+ \ldots \eeq
from which we read off the $NN\pi$ vertex in momentum space
\beq V_{NN\pi} = \frac{\krig{g}_A}{2F} \gamma_\mu q^\mu \gamma_5 \vec \tau
\quad , \eeq
where the momentum $q_\mu$ is out--going.
The transition amplitude for single pion emission off a nucleon takes the form
\beq T_{NN\pi} = -i\bar u (p') V_{NN\pi} u(p) = i
\frac{\krig{g}_A}{F}
\krig{m} \bar u (p') \gamma_5 u(p) \tau^i \, \, , \label{gtr1} \eeq
where I have used the Dirac equation $\bar u (p') \gamma_\mu q^\mu
\gamma_5 u(p) = -2 \krig{m} \bar u (p') \gamma_5 u(p)$ with $q
= p'-p$. Comparing eq.(\ref{gtr1}) with the canonical form of the
transition amplitude
\beq  T_{NN\pi} = i \krig{g}_{\pi N} \, \bar u (p') \gamma_5 u(p) \eeq
one arrives directly at the GTR,
\beq  \krig{g}_{\pi N} = \frac{\krig{g}_A  \krig{m}}{F} \, \, , \eeq
which is fulfilled within 5$\%$ in nature. This relation is
particularly intriguing because it links the strong pion--nucleon
coupling constant to some weak interaction quantities like $g_A$ and $F_\pi$
as a consequence of the chiral symmetry. Finally, if one wants to
discuss processes with  two (or more) nucleons in the initial and
final state, one has to add a string of terms of the type
\beq {\cal L}_{\bar \Psi \Psi \bar \Psi \Psi} +
{\cal L}_{\bar \Psi \Psi \bar \Psi \Psi \bar \Psi \Psi} + \dots \eeq
which are also subject to a chiral expansion and contain LECs which
can only be determined in few or many nucleon (baryon) processes.

Clearly, the appearance of the mass scale $\krig{m}$ in eq.(\ref{lpin}) causes
trouble. To be precise, if one calculates the self--energy of the
nucleon mass to one loop, one encounters terms of dimension {\it
  zero}, i.e. in dimensional regularization one finds a term of the
type \cite{GSS}
\beq {\cal L}_{\pi N}^{(0)} =
c_0 \, \bar \Psi \Psi \, , \quad \, c_0 \sim \left(\frac{\krig{m}}{F}\right)^2
\frac{1}{d-4} + \ldots \quad , \label{massdiv} \eeq
where the ellipsis stands for terms which are finite as $d \to 4$.
Such terms clearly make it difficult to organize the chiral expansion
in a straightforward and simple manner. They can only be avoided if
the additional mass scale $\krig{m} \sim 1$ GeV can be eliminated from the
lowest order effective Lagrangian. (Notice here the difference to the
pion case - there the mass vanishes as the quark masses are sent to
zero.) To do that, consider the mass of the nucleon large compared to
the typical external momenta transferred by pions or photons and write
the nucleon four--momentum as
\beq p_\mu = m \, v_\mu + \ell_\mu \, , \quad p^2 = m^2 \, , \quad v \cdot
\ell \ll m \, . \eeq
Notice that to this order we do not have to differentiate between $m$
and $\krig{m}$ and $v_\mu$ is the nucleon four--velocity (in the
rest--frame, we have $v_\mu =( 1 , \vec 0 \, )$). In that
case, we can decompose the wavefunction $\Psi$ into velocity
eigenstates \cite{JM} \cite{BKKM}
\beq \Psi (x) = \exp [ -i \krig{m} v \cdot x ] \, [ H(x) + h(x) ] \eeq
with
\beq P_v H = H \, , \, P_v h = -h \, , \quad P_v =
\frac{1}{2}(1+\gamma^\mu v_\mu) \, . \eeq
One now eliminates the 'small' component $h(x)$ either by using the
equations of motion or path--integral methods.
The Dirac equation for the velocity--dependent
baryon field $H = H_v$ (I will always suppress the label '$v$')
takes the form $i v \cdot \partial H_v = 0$ to lowest
order in $1/m$. This allows for a consistent chiral counting as described
below and the effective pion--nucleon Lagrangian takes the form:
\beq {\cal L}_{\pi N}^{(1)}  = \bar{H}
 \left( i v \cdot D  + \krig{g}_A S \cdot u \right) H
+ {\cal O}\left(\frac{1}{m} \right) \, , \label{lagr} \eeq
with $S_\mu$ the covariant spin--operator
\beq S_\mu = \frac{i}{2} \gamma_5 \sigma_{\mu \nu} v^\nu \, , \,
S \cdot v = 0 \, , \, \lbrace S_\mu , S_\nu \rbrace = \frac{1}{2} \left(
v_\mu v_\nu - g_{\mu \nu} \right) \, , \, [S_\mu , S_\nu] = i
\epsilon_{\mu \nu \gamma \delta} v^\gamma S^\delta \,
 \, , \label{spin}\eeq
in the convention $\epsilon^{0123} = -1$.
Two important observations can be made. Eq.(\ref{lagr}) does not
contain the nucleon mass term any more and also, all Dirac matrices
can be expressed as combinations of $v_\mu$ and $S_\mu$ \cite{JM},
\begin{displaymath} \bar H \gamma_\mu H = v_\mu \bar H H \, , \,
\bar H \gamma_5 H = 0 \, , \, \bar H \gamma_\mu \gamma_5 H =
2 \bar H S_\mu H \, \, ,  \end{displaymath}
\beq \bar H \sigma_{\mu \nu} H = 2 \epsilon_{\mu \nu \gamma
  \delta} v^\gamma \bar H S^\delta H \, , \, \bar H \gamma_5
\sigma_{\mu \nu} H = 2i \bar H (v_\mu S_\nu - v_\nu S_\mu) H \, \,
,\eeq
to leading order in $1/m$. More precisely,
this means e.g. $\bar H \gamma_5 H = {\cal O}(1/m)$.
 We read off the nucleon propagator,
\beq S_N (\omega ) = \frac{i}{\omega + i \eta} \, , \quad \omega = v
\cdot \ell \, , \quad \eta > 0\, \, , \label{prop} \eeq
and the Feynman insertion for the emission a pion with momentum $\ell$
from a nucleon is
\beq \frac{g_A}{F_\pi} \, \tau^a \, S \cdot \ell \quad . \label{vert}
\eeq
Notice that from now on I will  not always distinguish  between the
observables and their chiral limit values (although that distinction
should be kept in mind). Before proceeding with some actual
calculations in heavy baryon CHPT (HBCHPT),
let me outline the chiral power counting which is used
to organize the various terms in the energy expansion.

\section{Chiral power counting}
\label{sec:count}
To calculate any process to a given order, it is useful to have
a compact expression for the chiral power counting \cite{wein79} \cite{ecker}.
First, I will restrict myself to purely mesonic or single--baryon
processes. Since these arguments are general, I will consider the
three flavor case.
Any amplitude for a given physical process has a certain {\bf chiral} {\bf
dimension} $D$ which keeps track of the powers of external momenta and meson
masses. The building blocks to calculate this chiral dimension
from a general Feynman diagram in the CHPT loop expansion are
(i) $I_M$ Goldstone boson (meson) propagators $\sim 1/(\ell^2 -M^2)$
(with $M=M_{\pi , K, \eta}$ the meson mass) of dimension $D= -2$,
(ii) $I_B$ baryon propagators $\sim 1/ v \cdot \ell$ (in HBCHPT) with
$D= -1$, (iii) $N_d^M$ mesonic vertices with $d =2,4,6, \ldots$ and
(iv) $N_d^{MB}$ meson--baryon vertices with $d = 1,2,3, \ldots$.
Putting these together, the chiral dimension $D$ of a given amplitude reads
\begin{equation}
D =4L - 2I_M - I_B + \sum_d d( \,  N_d^M + N_d^{MB} \, )
\end{equation}
with $L$ the number of loops. For connected diagrams, one can use
the general topological relation
\begin{equation}
L = I_M + I_B -  \sum_d ( \,  N_d^M + N_d^{MB} \, ) + 1
\end{equation}
to eliminate $I_M$~:
\begin{equation}
D =2L + 2 + I_B + \sum_d (d-2)  N_d^M + \sum_d (d-2) N_d^{MB} ~ .
\label{DLgen}
\end{equation}
Lorentz invariance and chiral symmetry demand $d \ge 2$ for mesonic
interactions and thus the
term $\sum_d (d-2)  N_d^M$ is non--negative. Therefore, in the absence
of baryon fields, Eq.~(\ref{DLgen}) simplifies to \cite{wein79}
\begin{equation}
D =2L + 2 + \sum_d (d-2)  N_d^M  \, \ge 2L + 2 ~ .
\label{DLmeson}
\end{equation}
To lowest order $p^2$, one has to deal with tree diagrams
($L=0$) only. Loops are suppressed by powers of $p^{2L}$.

Another case of interest for us has a single baryon line running through
the diagram (i.e., there is exactly one baryon in the in-- and one baryon
in the out--state). In this case, the identity
\begin{equation}
\sum_d N_d^{MB} = I_B + 1
\end{equation}
holds leading to \cite{ecker}
\begin{equation}
D =2L + 1  + \sum_d (d-2)  N_d^M + \sum_d (d-1)  N_d^{MB}  \, \ge 2L + 1 ~ .
\label{DLMB}
\end{equation}
Therefore, tree diagrams start to contribute at order $p$ and one--loop
graphs at order $p^3$. Obviously, the relations involving
baryons are only valid in HBCHPT.

Let me now consider diagrams with $N_\gamma$ external photons.
Since gauge fields like the electromagnetic field appear
in covariant derivatives, their chiral dimension is obviously $D=1$.
One therefore writes the chiral dimension of a general amplitude
with $N_\gamma$ photons as
\begin{equation}
D = D_L + N_\gamma  ~,
\end{equation}
where $D_L$ is the degree of homogeneity of the (Feynman) amplitude $A$ as
a function of external momenta ($p$) and meson masses ($M$) in the
following sense (see also \cite{rho}):
\begin{equation}
A(p,M;C_i^r(\lambda),\lambda/M)
= M^{D_L} \, A ( p/M , 1;C_i^r(\lambda), \lambda/M )  ~ ,
\end{equation}
where $\lambda$ is an arbitrary renormalization scale and $C_i^r(\lambda)$
denote renormalized LECs. From now on, I will suppress the explicit dependence
on the renormalization scale and on the LECs. Since the total amplitude
is independent of the arbitrary scale $\lambda$, one may in particular
choose $\lambda = M$.
Note that $A(p,M)$ has also a certain physical dimension (which is
of course independent of the number of loops and is therefore in
general different from $D_L$). The correct
physical dimension is ensured by appropriate factors of $F_\pi$ and $m$
in the denominators.

Finally, consider a process with $E_n$ ($E_n = 4, 6, \ldots$) external
baryons (nucleons). The corresponding chiral dimension $D_n$ follows to be
\cite{weinnn}
\beq D_n = 2(L-C) + 4 - \frac{1}{2}E_n + \sum_i V_i \Delta_i \, ,
\quad \Delta_i = d_i + \frac{1}{2}n_i -2 \, \, , \eeq
where $C$ is  the number of connected pieces and one has $V_i$
vertices of type $i$ with $d_i$ derivatives and $n_i$ baryon fields
(these include the mesonic and meson-baryon vertices discussed before).
Chiral symmetry demands $\Delta_i \ge 0$. As before, loop diagrams
are suppressed by $p^{2L}$. Notice, however, that this chiral counting
only applies to the irreducible diagrams and not to the full S--matrix
since reducible diagrams can lead to IR pinch singularities and need
therefore a special treatment (for details, see refs.\cite{weinnn}
\cite{weinnp}).

I will now briefly discuss the general structure of the effective
Lagrangian based on these power counting rules, restricting myself
again to the two--flavor case and processes with one nucleon in the
asymptotic in-- and out--states. While the lowest order Lagrangian
eq.(\ref{lagr}) has $D=1$, one can construct a string of local
operators with $D=2,3,4, \ldots$. One--loop diagrams start at order
$p^3$ if one only uses insertions from ${\cal L}_{\pi
  N}^{(1)}$. Two--loop graphs are suppressed by two more powers of $p$
so that within the one--loop approximation one should consider tree
diagrams from
\beq
{\cal L}_{\pi N} = {\cal L}_{\pi N}^{(1)} + {\cal L}_{\pi N}^{(2)} +
{\cal L}_{\pi N}^{(3)} + {\cal L}_{\pi N}^{(4)}  \, \, , \label{lpin4} \eeq
and loop diagrams with insertions from ${\cal L}_{\pi N}^{(1,2)}$.
 It is important to stress that not all of the terms in
${\cal L}_{\pi N}^{(2,3,4)}$ contain LECs, but some of  the
coefficients are indeed fixed for kinematical or similar reasons. I
will discuss the general structure of ${\cal L}_{\pi N}^{(2)}$ in
section~\ref{sec:lpin2}.
It should also be stressed that although there are many terms in
${\cal L}_{\pi N}^{(2,3,4)}$, for a given process most of them do {\it
  not} contribute. As an example let me quote the order $q^4$
calculation of the nucleons' electromagnetic polarizabilities
\cite{bkms} which involves altogether four LECs from ${\cal L}_{\pi N}^{(2)}$
and four from ${\cal L}_{\pi N}^{(4)}$, a number which can certainly be
controlled. To all of this, one has of course to add the purely
mesonic Lagrangian ${\cal L}_{\pi \pi}^{(2)}+{\cal L}_{\pi \pi}^{(4)}$
\cite{GL1} \cite{GL2}.

\section{A simple calculation}
\label{sec:calc}
In this section, I will present a typical calculation, namely the
nucleon mass shift from the pion loop (in the one--loop
approximation). The full result can e.g. be found in refs.\cite{dobo}
\cite{BKKM}. A similar sample calculation has been given in the
lectures by Jenkins and Manohar \cite{dobo}, but I will use another
method which is easier to generalize to processes with external photons.

Consider the Feynman diagram where the nucleon emits a pion of momentum
$\ell$ and absorbs the same pion (which is incoming with momentum
$-\ell$). Using eqs.(\ref{prop},\ref{vert}) and the relativistic
propagator for the pion, the mass shift $\delta m$ is given by
\beq
\delta m = i \frac{3g_A^2}{F_\pi^2} \int \frac{d^d \ell}{(2\pi)^d}
\frac{i}{\ell^2 - M_\pi^2 + i\eta} \,  \frac{i}{-v \cdot \ell + i \eta}
 S \cdot (- \ell) \,  S \cdot \ell \, , \eeq
making use of $\tau^a \tau^a = 3$. From the anti--commutation relation
of two spin matrices, eq.(\ref{spin}), and by completing the square we have
\beq S_\mu S_\nu \ell^\mu \ell^\nu = \frac{1}{4} \left( v \cdot \ell
\, v \cdot \ell + M_\pi^2 - \ell^2 - M_\pi^2 \right) \, , \eeq
so that
\beq
\delta m = i \frac{3g_A^2}{4 F_\pi^2} \int \frac{d^d \ell}{(2\pi)^d} \left[
\frac{1}{v \cdot \ell -i \eta} +
\frac{v \cdot \ell}{M_\pi^2 - \ell^2 -i\eta} - \frac{M_\pi^2}{(M_\pi^2
  - \ell^2 -i\eta)(v \cdot \ell - i\eta)} \right] \, . \label{int} \eeq
To calculate this integral, we make use of dimensional
regularization. The first term in the square brackets vanishes in
dimensional regularization (see e.g. ref.\cite{coll}) and the second
one is odd under $\ell \to -\ell$, i.e. it also vanishes. So we are
left with
\beq \delta m =  \frac{3g_A^2}{4 F_\pi^2} \, J(0) \, M_\pi^2 \eeq
\beq J(0) = \frac{1}{i} \, \int \frac{d^d \ell}{(2\pi)^d}
\frac{1}{(M_\pi^2  - \ell^2 -i\eta)(v \cdot \ell - i\eta)} \quad . \eeq
The remaining task is to evaluate $J(0)$. For that, we use the
identity
\beq \frac{1}{AB} = \int_0^\infty dy \, \frac{2}{[A+2yB]^2} \, \, , \eeq
define $\ell' = \ell - y v$, complete the square and use $v^2 =1$,
\beq J(0)  = \frac{1}{i} \, \int_0^\infty dy \, \int \frac{d^d
\ell '}{(2\pi)^d} \, \frac{1}{[M_\pi^2 + y^2 + {\ell'}^2 - i \eta]^2}
\eeq
\beq \quad \quad \quad \quad \quad  =
\frac{2}{(2\pi)^d} \, \int_0^\infty dy \, \int_0^\infty d\ell'
\frac{(\ell')^{d-1}}{[M_\pi^2 + y^2 + {\ell'}^2]^2} \,
\frac{2 \pi^{d/2}}{\Gamma(d/2)} \, \, , \label{j0} \eeq
where we have performed a Wick rotation,
$\ell_0 \to i \ell_0$ and dropped the $i
\eta$. The last factor in eq.(\ref{j0}) is the surface of the sphere
in $d$ dimensions. Introducing polar coordinates, $y = r \cos \phi$ and
$\ell' = r \sin \phi$ and noting that the Jacobian of this
transformation  is $r$, we have
\beq J(0) = \frac{4(4 \pi)^{-d/2}}{\Gamma(d/2)} \, \int_0^\infty dr \,
\frac{r^d}{(r^2+M_\pi^2)^2} \,  \int_0^{\pi/2} d\phi \, (\sin
\phi)^{d-1} \, . \label{ja} \eeq
We then perform the further substitution $r = M_\pi \tan \Phi$ in the
$r$--integral. Then,
both integrals appearing in eq.(\ref{ja}) can be expressed in terms
of products of $\Gamma$ functions with the result
\beq J(0) = M_\pi^{d-3} \, (4 \pi)^{-d/2} \, \Gamma \left(
\frac{1}{2} \right) \, \Gamma \left(
\frac{3-d}{2} \right) = - \frac{M_\pi}{8 \pi} \, \, , \eeq
where in the last step I have set $d=4$. This leads us to
\beq \delta m = i \, \Sigma = -\frac{3g_A^2}{32 \pi} \,
\frac{M_\pi^3}{F_\pi^2} \, \, . \label{self} \eeq
The pion loop leads to a self--energy $\Sigma$ which shifts the pole of
the nucleon propagator by $\delta m$, i.e.
\beq \frac{i}{v\cdot \ell - i \Sigma} = \frac{i}{v \cdot \ell} \left(
1 - \Sigma \frac{i}{v \cdot \ell} \right)^{-1} = \frac{i}{v\cdot
  \ell}+ \frac{i}{v\cdot \ell} \, \Sigma \, \frac{i}{v\cdot \ell} + \ldots =
 \frac{i}{v \cdot \ell - \delta m} \, . \eeq
There are a few important remarks concerning eq.(\ref{self}). First,
the pion loop contribution is non--analytic in the quark masses
\cite{lapa}
\beq \delta m \sim (\hat m)^{3/2}    \, \, , \eeq
 since
\beq
M^2_\pi = B \, \hat m \, [1+{\cal O}(m_{quark})]~.
\label{masses}
\eeq
The constant $B$ is related to the scalar quark condensate and is
assumed to be non--vanishing in the chiral limit (supported by lattice data).
(For a different scenario, see e.g. refs.\cite{Stern}). Second, the
pion cloud contribution is attractive, i.e. it lowers the nucleon
mass, and third, it vanishes in the chiral limit, i.e it has the
expected chiral dimension of three (since it is a one--loop graph with
insertions from the lowest order effective Lagrangian). More detailed
studies of the baryon masses
and $\sigma$--terms can be found in refs.\cite{gass} \cite{dobo}
\cite{liz} \cite{jms} \cite{bkmz} \cite{ll} \cite{samir} \cite{mary}.
Here, I only wish to
state that at present it is not known whether the scalar three--flavour
sector (i.e. the baryon masses and the $\sigma$--terms) can be described
consistently within CHPT. It seems that one at least should go to
order $q^4$ and it is unclear how the close--by spin--3/2 decuplet
has to be treated (i.e. as effective degrees of freedom or frozen to
supply large contributions to certain LECs). The problem is that the
kaon loop corrections are rather large (since $(M_K / M_\pi)^3 \sim 46$)
and the corresponding LECs have therefore large coefficients to
compensate for the meson cloud contribution. It might, however, well be that
the corrections beyond leading order are much smaller. This can
only be decided upon a {\it complete} calculation to order $q^4$.

\section{The structure of $\boldmath{{\cal L}_{\pi N}^{(2)}}$}
\label{sec:lpin2}
In this section, I will first write down the order $p^2$ effective
Lagrangian, ${\cal L}_{\pi N}^{(2)}$, and then discuss some of its
peculiarities. Allowing for the moment for $m_u \ne m_d$, its
most general form is (I only consider external scalar and vector
fields, the generalization to pseudoscalar and axial--vector ones is
straightforward):
\begin{displaymath}
{\cal L}_{\pi N}^{(2)} = \bar H \left\lbrace
\frac{1}{2\krig{m}}(v \cdot D)^2 - \frac{1}{2\krig{m}} D^2
-\frac{i \krig{g}_A}{2 \krig{m}} \lbrace S \cdot D , v \cdot u \rbrace
+ c_1 \, {\rm Tr} \, \chi_+ +
 \left(c_2 - \frac{\krig{g}_A^2}{8 \krig{m}} \right)
(v \cdot u)^2 \right. \end{displaymath}
\beq
 \left.
+ c_3 \, u \cdot u + \left( c_4 + \frac{1}{4\krig{m}} \right) [S^\mu,S^\nu]
u_\mu u_\nu +  c_5 \, \tilde{\chi}_+ -  \frac{i [S^\mu,S^\nu]}{4 \krig{m}}
\left( (1 +c_6) f_{\mu \nu}^+ + c_7 \, {\rm Tr} \,  f_{\mu \nu}^+ \right)
 \right\rbrace H  \label{lp2}\eeq
with
\beq \chi_\pm = u^\dagger \chi u^\dagger \pm u \chi^\dagger u \, , \quad
\tilde{\chi}_+ = \chi_+ - \frac{1}{2}{\rm Tr}\, (\chi_+) \, , \quad
f_{\mu \nu}^+ = u^\dagger F_{\mu \nu} u + u F_{\mu \nu} u^\dagger \, .
 \eeq
Here, $\chi = 2 B {\cal M}$ (${\cal M}$ is the quark mass matrix) and
$F_{\mu \nu} = \partial_\mu A_\nu - \partial_\nu A_\mu$ the canonical
photon field strength tensor.
The term proportional to $c_5$ vanishes in the isospin limit $m_u= m_d$.
One observes that some of the terms in eq.({\ref{lp2}) have no LECs but
rather fixed coefficients. The origin of this is clear, these terms
stem from the expansion of the relativistic $\pi N$ Lagrangian. To see
this in more detail, use the equation of motion for the small
component field $h(x)$,
\beq h = \frac{1}{2} (1 - \barr{v}) \frac{1}{2 \krig{m}} \left( i
\barr{D} +\frac{\krig{g}_A}{2} \barr{u} \gamma_5 \right) H + {\cal
  O}(1/ \krig{m}^2 )  \, \,  \eeq
to construct
\beq {\cal L}_{\pi N}^{(2)} = \frac{1}{2 \krig{m}} \bar H ( i
\barr{D} +\frac{\krig{g}_A}{2} \barr{u} \gamma_5 )
\frac{1 -  \barr{v}}{2} \bar H (i \barr{D} +\frac{\krig{g}_A}{2} \barr{u}
\gamma_5 ) H \, \, .  \label{lp2r} \eeq
Altogther, we have four different products of terms in eq.(\ref{lp2r}). Let me
consider the one proportional to $\barr{D} \barr{D}$, the other
contributions can be calculated in a similar fashion:
\beq \frac{i^2}{2 \krig{m}} \bar H \barr{D} \frac{1 -  \barr{v}}{2}
\barr{D} H = -\frac{1}{2 \krig{m}}D^\mu D^\nu \bar H \left[ \gamma_\mu
\frac{1 -  \barr{v}}{2} \gamma_\nu \right] H \, \, . \eeq
Straightforward application of the $\gamma$--matrix algebra allows us
to write the term in the square brackets on the r.h.s. as
\beq \bar H[ \dots ]H = \bar H [
g_{\mu \nu} - i \sigma_{\mu \nu} - \gamma_\mu v_\nu ] H \, \, , \eeq
where I have used that $P_v H = H$. Collecting pieces, we have
\beq \frac{i^2}{2 \krig{m}} \bar H \left\lbrace  g_{\mu \nu} D^\mu
D^\nu  -  2 i \epsilon_{\mu \nu \alpha \beta}  D^\mu
D^\nu v^\alpha S^\beta  -  v_\mu D^\mu v_\nu D^\nu  \right\rbrace  H\, \,
. \eeq
This simplifies further since $\epsilon_{\mu \nu \alpha \beta}$  and
$D^\mu D^\nu$ are anti-- and symmetric under $\mu \leftrightarrow
\nu$, respectively,
\beq \frac{1}{2 \krig{m}} \bar H \left\lbrace  -D^2 + (v \cdot D)^2 +
i \epsilon_{\mu \nu \alpha \beta} [D^\mu , D^\nu] v^\alpha S^\beta
\right \rbrace H \, \, . \eeq
Finally, the commutator of two chiral covariant derivatives is
related to the chiral connection $\Gamma_\mu$ via
\beq [ D_\mu , D_\nu] = \partial_\mu \Gamma_\nu - \partial_\nu
\Gamma_\mu +  [\Gamma_\mu , \Gamma_\nu ]  \, \, \, . \eeq
We can work this out for the explicit form of the connection given in
eq.(\ref{conn}) and find
\beq [ D_\mu , D_\nu] = -\frac{i}{2} f_{\mu \nu}^+ + \frac{1}{4}
[u_\mu , u_\nu] \quad . \eeq
Putting everything together, we find that the first three terms in
eq.(\ref{lp2}) plus the piece proportional to $\krig{g}_A^2 / (8
\krig{m})$ in the fifth and the piece proportional to $1/( 4 \krig{m})$ in
the seventh term are generated by expanding the operator $\barr{D} \barr{D}$.
 The first two contributions are corrections to the
kinetic energy and contain (besides others) a two--photon--nucleon
seagull which leads to the correct LET for low--energy Compton
scattering (see section \ref{sec:LET}). Similarly, the third term has
no free coefficient since it gives the leading term in the quark mass
expansion of the electric dipole amplitude $E_{0+}$ in neutral pion
photoproduction off protons, $E_{0+} \sim M_\pi / m$. These terms have
no direct relativistic counter parts but are simply due to the
expansion of the relativistic pion--nucleon effective Lagrangian.
Such constraints were
e.g. not accounted for in eq.(17) of ref.\cite{JM}. The finite
coefficients $c_{1,2,3,4}$ can be determined from the pion--nucleon
$\sigma$--term and S-- and P--wave $\pi N$ scattering lengths
\cite{BKKM} \cite{bkmpin} \cite{bkmrev}. The last two terms in eq.(\ref{lp2})
are easy to pin down, they are related to the isoscalar and isovector
anomalous magnetic moment of the nucleon (in the chiral limit)
\cite{BKKM}
\beq c_6 = \krig{\kappa}_V \, \, \quad c_7 = \frac{1}{2}
(\krig{\kappa}_S - \krig{\kappa}_V ) \quad . \eeq
Apart from the LEC $c_5$, all of these coefficients have been
determined. For a meaningful comparison, I normalize all terms to $(1 /
2 \krig{m})$, i.e. I define
\beq c_i' = 2 \krig{m} c_i \, \, (i=1,2,3,4) \, , \quad c_{6,7}' = 2
c_{6,7} \, \, \, , \eeq
and use $\krig{m} = m= (m_p +m_n)/2 = 0.93892$ GeV (which is correct to
this order). The resulting values are summarized in table 1.
\bigskip

\begin{center}
\begin{tabular}{|crll|} \hline
LEC    & Value  \quad \quad & Source             & Ref.       \\ \hline
$c_1'$   & $-1.63\pm 0.21$  & $\sigma_{\pi N}(0)$& \cite{BKKM} \\
$c_2'$   & $6.20\pm 0.38$  & $\pi N \to \pi N  $& \cite{bkmpin}
\cite{bkmrev} \\
$c_3'$   & $-9.86\pm 0.41$  & $\pi N \to \pi N  $& \cite{bkmpin}
\cite{bkmrev} \\
$c_4'$   & $7.73\pm 0.18$  & $\pi N \to \pi N  $& \cite{bkmrev} \\
$c_6'$   & $7.41\pm 0.00$  & $\kappa_{p,n}     $& \cite{BKKM} \\
$c_7'$   & $-7.17\pm 0.00$ & $\kappa_{p,n}     $& \cite{BKKM}
 \\ \hline \end{tabular}
\smallskip

Table 1: Numerical values of the LECs in ${\cal L}_{\pi N}^{(2)}$
(with the exception of $c_5$).

\end{center}

\noindent I would like to stress that the
coefficients $c_1$ and $c_2$ have been
determined from the chiral corrections to  the $\pi N$ $\sigma$--term
and to the S--wave scattering length $a^+$. Both of the quantities are
either subject to large higher order corrections or are not very
accurately determined empirically. For the LEC $c_3'$, there is
another way of fixing it. The value given in the table stems from the
so--called axial polarizability $\alpha_A$ and the error solely
reflects the uncertainty in the empirical value of $\alpha_A$. One can
also use the pion--nucleon P--wave scattering volume $P_1^+ \sim
(4a_{33}+2a_{31}+2a_{13} +a_{11})$
(with $a_{2I,2J}$ the scattering volumes in the respective channels) to get
this LEC. One finds $c_3' = -11.87 \pm 2.69$, which is somewhat larger
than the value given in table 1 but overlaps within one standard
deviation.  If one is conservative, one would therefore estimate
the theoretical error to be $\pm 1.5$ for $c_i'$ $(i=1,2,3,4)$. For
$c_6'$ and $c_7'$, this theoretical uncertainty is clearly much
smaller since the anomalous magnetic moments of the proton and the
neutron are well known. In any case, it would  be very useful
to have another and possibly more accurate determination of these LECs.
In fact, van Kolck and collaborators \cite{bira}
have considered the chiral expansion of the nucleon--nucleon
interaction to one--loop accuracy. Naturally, they also have some of
the terms appearing in eq.(\ref{lp2}) and they call the corresponding
LECs $B_{1,2,3}$. Translating their language into the one used here,
one has
\beq B_1 = 4 c_3 \, , \quad B_2 = 8 c_1 \, , \quad B_3 =-4c_4 -
 \frac{1}{\krig{m}} \quad . \eeq
In a best fit to the deuteron properties and the NN phase shifts at
low energies these parameters were left completely unconstrained. The
resulting values are $B_1 = 3.42$, $B_2 = 8.55$ and $B_3 = 17.7$ \cite{birat}
 which translates into $c_1' = 2.01$, $c_3' = 1.60$ and $c_4' = -8.80$
(all in GeV$^{-1}$). The value for $c_2'$ is considerably smaller
than the one given in table~1 whereas $c_1'$ and $c_4'$ are of
comparable magnitude but have the
opposite sign to the numbers given in table~1. To my opinion, this
discrepancy is rooted in the fact that the fit contains a rather large
number of LECs (mostly related to four--nucleon contact terms) so that
the individual ones like $B_{1,2,3}$ are not very precisely determined
(lowest order one--pion exchange and some of the contact terms are
most important to give the gross features of the NN interaction, so
that a term like the one $\sim B_3$, which is the $\sigma$--term, does
not carry too much statistical weight). In fact, one should rather use
the values given from $\pi N$ scattering as input in the much tougher
problem of the nuclear forces. Naturally, one can ask the question
whether one can understand the values of the $c_i'$ given in table~1?
This is indeed the case as detailed in ref.\cite{bkmrev}.
To be specific, let me pick $c_3'$.
Consider an EFT of meson resonances (M =
V,A,S,P) chirally coupled to the Goldstone bosons and matter fields
($N$) as well as baryonic excitations ($N^\star$). Integrating out the
meson and nucleon resonances,
\beq \int [dM][dN^\star] \exp i \int dx \tilde{{\cal L}}_{\rm eff}
 [U,M,N,N^\star] =  \exp i \int dx {\cal L}_{\rm eff} [U,N]  \, \, , \eeq
one is left with a string of higher dimensional operators
($D= 2,3,\ldots$) contributing to the effective pion--nucleon
Lagrangian in a manifestly chirally invariant manner and with
coefficients given entirely in terms of resonance masses and coupling
constants of the resonance fields to the Goldstone bosons and
nucleons. In the meson sector, this works remarkably well \cite{reso}. Here,
it has more the status of a working hypothesis (but a well--founded one).
For the LEC $c_3'$, we have $\Delta$, $N^\star (1440)$ and scalar
meson exchange.  Varying the corresponding couplings within their
allowed values, one finds \cite{ulfmit} \cite{bkmrev}
\beqa & c_{3, \rm res}'  = c_{3, \Delta}' + c_{3, N^\star}' + c_{3, S}'
\nonumber \\
& = (-4.7 \ldots -6.0) + (-0.2 \ldots -0.4) + (-1.9 \ldots -3.0)
\nonumber \\
& = -6.8 \ldots -9.4 \quad . \eeqa
This number is of comparable size to the one given in table~1,
although the resulting values tend to come out on the lower side.
Nevertheless, such considerations seem to give credit to the
resonance saturation picture of the LECs also in the baryon sector.  It is
worth to point out that besides a large contribution from the $\Delta$,
scalar meson exchange gives a sizeable chunk to $c_{3, {\rm
  res}}'$. This shows that the resonance exchange picture is more
complex in the baryon sector than in the mesonic one. More work in
this direction is clearly needed.

\medskip

The important lesson to be learned from this discussion is that in
HBCHPT we find in ${\cal L}_{\pi N}^{(2,3, \dots)}$ terms which have
no LECs but rather coefficients which are fixed. This is an artefact
of the {\it dual } expansion in small momenta $p$ versus the chiral
symmetry breaking scale {\it and} versus inverse powers
of the nucleon mass, i.e.
\beq \frac{p}{\Lambda_\chi } \, \, , \quad \frac{p}{m} \, \, . \eeq
In practice, since $m \sim \Lambda_\chi$, one has essentially one
expansion parameter besides the one due to the effect of the finite
quark masses. Since loops only appear at $D=3$, all the LECs in
${\cal L}_{\pi N}^{(2)}$ are {\it finite} (as already should have
become clear from the previous discussion). At the next order, $p^3$,
divergences appear. Ecker \cite{eckp3} has recently calculated the
full determinant using heat--kernel methods and given the divergent
terms,
\beq {\cal L}_{\pi N}^{(3)} = \frac{1}{(4\pi)^2} \, \sum_{i=1}^{22} \,
b_i \, \bar H (x) \, O_i (x) \, H(x)       \eeq
with
\beq
b_i = b_i^r (\lambda) + \Gamma_i \, L(\lambda)
\, \, , \quad L \sim \frac{1}{d-4} \quad . \eeq
The $O_i$ are monomials in the fields and have dimension three. Their
explicit forms together with the values of the $\Gamma_i$ can be found
in ref.\cite{eckp3}. Only a few of the finite $b_i^r$ have either been
determined  from phenomenology or estimated from resonance exchange
\cite{liz} \cite{BKKM} \cite{bkmpin} \cite{bkmpi0}. For
$  {\cal L}_{\pi N}^{(4)}$, no systematic study exists so far but some
dimension four operators have been used and their corresponding  LECs
fixed \cite{bkms} \cite{bkmpi0}.

\section{The meaning of low--energy theorems (LETs)}
\label{sec:LET}

In this section, I will briefly discuss the meaning of the so--called
low--energy theorems. More details are given in ref.\cite{gerulf}.
Let us first consider a well--known example of a LET
 involving the electromagnetic current. Consider the
scattering of very soft photons on the proton, i.e., the Compton scattering
process $\gamma (k_1) + p(p_1) \to \gamma (k_2) + p(p_2)$
and denote by $\ve \, (\ve ')$
the polarization vector of the incoming (outgoing) photon. The transition
matrix element $T$ (normalized to $d\sigma / d\Omega = |T|^2$) can be
expanded in a Taylor series in the small parameter $\delta =
|\vec{k_1}|/m$.
In the forward  direction and in a gauge where
the polarization vectors have only space components, $T$ takes the form
\begin{equation}
T = c_0 \, \vec{\ve}\, ' \cdot \vec{\ve} + i \, c_1 \, \delta
\, \vec{\sigma} \cdot (
\vec{\ve}\, ' \times \vec{\ve} \, ) + {\cal O}(\delta^2) ~ .
\label{Comp}
\end{equation}
The parameter $\delta$ can be made arbitrarily small in the laboratory so
that the first two terms in the Taylor expansion (\ref{Comp}) dominate.
To be precise, the first one proportional to $c_0$ gives the low--energy
limit for the spin--averaged Compton amplitude, while the second ($\sim c_1$)
is of pure spin--flip type and can directly be detected in polarized
photon proton scattering (to my knowledge, such a test has not yet
been performed). The pertinent LETs fix the values of $c_0$ and
$c_1$ in terms of measurable quantities \cite{low},
\begin{equation}
c_0 = - \frac{Z^2e^2}{4 \pi m} \, , \quad c_1 =
- \frac{Z^2e^2 \kappa_p^2}{8 \pi m}
\label{c01}
\end{equation}
with $Z =1$ the  charge of the proton. To arrive at Eq.~(\ref{c01}),
one only makes use of gauge
invariance and the fact that the $T$--matrix can be written in terms of a
time--ordered product of two conserved vector currents sandwiched between
proton states. The derivation proceeds by showing that for small
enough photon energies the matrix element is determined by the electromagnetic
form factor of the proton at $q^2 = 0$ \cite{low}.

Similar methods can be applied to other than the electromagnetic
currents. In strong interaction physics, a special role is played by the
axial--vector currents. The associated symmetries are spontaneously
broken giving rise to the Goldstone matrix elements
\beq
\langle 0|A^a_\mu(0)|\pi^b(p)\rangle =
i \delta^{ab} F_\pi p_\mu
\label{Gme}
\eeq
where $a,b$ are isospin indices. In the chiral limit,
the massless pions play a similar role as the photon and many
LETs have been derived for ``soft pions".
In light of the previous discussion on Compton
scattering, the most obvious one is Weinberg's prediction for elastic
$\pi p$ scattering \cite{wein68}. We only need the following translations~:
\begin{equation}
<p| T \, j_\mu^{\rm em} (x) j_\nu^{\rm em} (0)|p> \, \, \to \, \,
<p| T \, A_\mu^{\pi^+} (x) A_\nu^{\pi^-} (0)|p> ~,
\end{equation}
\begin{equation}
\partial^\mu j_\mu^{\rm em} = 0 \, \, \to \, \,
\partial^\mu A_\mu^{\pi^-}  = 0  ~.
\end{equation}
In contrast to photons, pions are not massless in the real
world. It is therefore interesting to find out how the LETs for
soft pions are modified in the presence of non--zero pion masses
(due to non--vanishing quark masses). In the old days of current
algebra, a lot of emphasis was put on the PCAC
(Partial Conservation of the Axial--Vector Current) relation,
consistent with the Goldstone matrix element (\ref{Gme}),
\begin{equation}
\partial^\mu A^a_\mu = M_\pi^2 F_\pi \pi^a ~ \quad .
\label{PCAC}
\end{equation}
Although the precise meaning of (\ref{PCAC}) has
long been understood \cite{col}, it does not offer a systematic
method to calculate higher orders in the momentum and mass expansion
of LETs. The derivation of non--leading terms in the days of
current algebra and PCAC was more an art than a science, often
involving dangerous procedures like off--shell
extrapolations of amplitudes. In the modern language, i.e. the EFT of
the Standard Model, these higher order corrections can be calculated
unambigously and one correspondingly defines a low--energy theorem via:

\vfill \eject

\begin{displaymath}
 {\rm{\bf L}(OW)} \quad {\rm {\bf E}(NERGY)} \quad{\rm  {\bf
 T}(HEOREM)} \quad {\rm OF} \quad  {\cal O}(p^n) \end{displaymath}
\beq \equiv {\rm GENERAL} \quad {\rm PREDICTION} \quad
 {\rm OF} \quad {\rm CHPT} \quad {\rm TO} \quad {\cal O}(p^n)  \quad . \eeq
\noindent
By general prediction I mean a
strict consequence of the SM depending on some LECs like
$F_\pi, m, g_A, \kappa_p, \ldots$, but without any model assumption for these
parameters. This definition contains a
precise prescription how to obtain higher--order corrections to
leading--order LETs.

The soft--photon theorems, e.g., for Compton scattering \cite{low},
involve the limit of small photon momenta, with all other momenta
remaining fixed. Therefore, they hold to all orders in the non--photonic
momenta and masses. In the low--energy expansion of CHPT, on the
other hand, the ratios of all small momenta and pseudoscalar meson
masses are held fixed. Of course, the soft--photon theorems are also
valid in CHPT as in any gauge invariant quantum field theory.
To understand this difference  of low--energy limits,
I will now rederive and extend the LET for spin--averaged nucleon
Compton scattering in the framework of HBCHPT \cite{BKKM}. Consider the
spin--averaged Compton amplitude in forward direction (in the Coulomb
gauge $\ve \cdot v = 0$)
\beq
e^2 \ve^\mu \ve^\nu
\frac{1}{4} {\rm Tr} \biggl[ (1 + \gamma_\lambda v^\lambda) T_{\mu \nu} (v,k)
\biggr] = e^2 \biggl[ \ve^2 U(\omega ) + (\ve \cdot k )^2
V(\omega ) \biggr]
\eeq
with $\omega = v \cdot k$ ($k$ is the photon momentum) and
\beq
T_{\mu \nu} (v,k) = \int d^4 k \, {\rm e}^{ik \cdot x} \, < N(v)| T j^{\rm
em}_\mu (x) j_\nu^{\rm em} (0) |N(v)>~.
\eeq
All dynamical information is
contained in the functions $U(\omega)$ and $V (\omega )$. We
only consider $U(\omega)$ here and refer to Ref.~\cite{BKKM} for
the calculation of both $U(\omega)$ and $V(\omega)$. In the
Thomson limit, only $U(0)$ contributes to the amplitude.
In the forward direction, the only quantities with non--zero chiral
dimension are $\omega$ and $M_\pi$. In order to make this dependence
explicit, we write $U(\omega,M_\pi)$ instead of $U(\omega)$. With
$N_\gamma = 2$ external photons, the degree of homogeneity $D_L$
for a given CHPT contribution to $U(\omega,M_\pi)$ follows from
Eq.~(\ref{DLMB})~:
\beq
D_L =2L - 1  + \sum_d (d-2)  N_d^M + \sum_d (d-1)  N_d^{MB}  \, \ge - 1 ~ .
\label{DLC}
\eeq
Therefore, the chiral expansion of $U(\omega,M_\pi)$ takes the following
general form~:
\beq
U(\omega,M_\pi) = \sum_{D_L\ge -1} \omega^{D_L} f_{D_L}(\omega/M_\pi)~.
\label{Uce}
\eeq
The following arguments illuminate the difference and the
interplay between the soft--photon limit and the low--energy expansion
of CHPT. Let us consider first the leading terms in the chiral expansion
(\ref{Uce})~:
\beq
U(\omega,M_\pi) = {1\over \omega} f_{-1}(\omega/M_\pi) +
 f_0(\omega/M_\pi) + {\cal O}(p^3)~.
\eeq
Eq.~(\ref{DLC}) tells us that only tree diagrams can contribute
to the first two terms. However, the relevant tree diagrams
do not contain pion lines. Consequently, the functions
$f_{-1}$, $f_0$ cannot depend on $M_\pi$ and are therefore constants.
Since the soft--photon theorem \cite{low} requires $U(0,M_\pi)$
to be finite, $f_{-1}$ must actually vanish and the chiral
expansion of $U(\omega,M_\pi)$ can be written as
\beq
U(\omega,M_\pi) = f_0 + \sum_{D_L\ge 1} \omega^{D_L} f_{D_L}(\omega/M_\pi)~.
\label{Uce2}
\eeq
But the soft--photon theorem yields additional information~:
since the Compton amplitude is independent of $M_\pi$ in the Thomson
limit and since there is no term linear in $\omega$ in the spin--averaged
amplitude, we find
\beq
\lim_{\omega \to 0}~ \omega^{n-1} f_n(\omega/M_\pi) = 0 \qquad
(n\ge 1) \label{spl}
\eeq
implying in particular that the constant $f_0$ describes the
Thomson limit~:
\beq
U(0,M_\pi) = f_0~.
\eeq
Let me now verify these results by explicit calculation.
In the Coulomb gauge, there is no direct
photon--nucleon coupling from the lowest--order effective Lagrangian
${\cal L}_{\pi N}^{(1)}$ since it is proportional to $\ve \cdot v$.
Consequently, the corresponding Born diagrams  vanish so that indeed
$f_{-1}=0$. On the other hand, I had argued in section \ref{sec:lpin2} that
the heavy mass expansion of the relativistic $\pi N$ Dirac Lagrangian
leads to a Feynman insertion of the form (from the first two terms in
eq.(\ref{lp2})):
\beq
i \frac{e^2}{m} \frac{1}{2} (1 + \tau_3 ) \biggl[ \ve^2 -( \ve
\cdot v)^2 \biggr] = i \frac{e^2 Z^2}{m} \,  \ve^2
\eeq
producing the desired result $f_0 = Z^2 / m$, the Thomson limit.

At the next order in the chiral expansion, ${\cal O}(p^3)$ ($D_L = 1$),
the function $f_1(\omega/M_\pi)$ is given by the
finite sum of 9 one--loop diagrams \cite{BKM} \cite{BKKM}. According
to Eq.~(\ref{spl}), $f_1$ vanishes for $\omega \to 0$. The term linear
in $\omega/M_\pi$ yields the leading contribution to the sum of the
electric and magnetic polarizabilities of the nucleon, defined by
the second--order Taylor coefficient in the expansion of $U(\omega,M_\pi)$
in $\omega$~:
\beq
f_1(\omega/M_\pi) = - {11 g_A^2 \omega\over 192 \pi F_\pi^2 M_\pi}
+ {\cal O}(\omega^2)~.
\eeq
The $1/M_\pi$ behaviour should not come as a surprise
-- in the chiral limit the  pion cloud becomes long--ranged (instead of being
Yukawa--suppressed) so that the polarizabilities explode.
This behaviour is specific to the leading contribution
of ${\cal O}(p^3)$. In fact, from the general form (\ref{Uce2}) one
immediately derives that the contribution of ${\cal O}(p^n)$
($D_L = n - 2$) to the polarizabilities is of the form $c_n M_\pi^{n-4}$
($n\ge 3$), where $c_n$ is a constant that may be zero.
One can perform a similar analysis for the amplitude
$V(\omega)$ and for the spin--flip amplitude. We do not
discuss these amplitudes here but refer the reader to Ref.~\cite{BKKM}
for details.

Next, let us consider the processes
$\gamma N \to \pi^0 N$ $(N=p,n)$
at threshold, i.e., for vanishing three--momentum of the pion
in the nucleon rest frame. At
threshold, only the electric dipole amplitude $E_{0+}$ survives and
the only quantity with non--zero chiral dimension is $M_\pi$.
In the usual conventions,
$E_{0+}$ has physical dimension $-1$ and it can therefore be
written as
\beq
E_{0+} = {e g_A \over F}~A\left( {M_\pi\over m},{M_\pi\over F}\right)~.
\eeq
The dimensionless amplitude $A$ will be expressed as a power
series in $M_\pi$. The various parts are characterized by the
degree of homogeneity (in $M_\pi$) $D_L$ according to the chiral
expansion. Since $N_\gamma=1$ in the present case, we obtain from
Eq.~(\ref{DLMB})
\beq
D_L = D-1 = 2L + \sum_d (d-2)  N_d^M + \sum_d (d-1)  N_d^{MB} ~ .
\eeq
For the LET of ${\cal O}(p^3)$ in question, only lowest--order mesonic vertices
($d=2$) will appear. Therefore, in this case the general formula for $D_L$
takes the simpler form
\beq
D_L = 2L + \sum_d (d-1)  N_d^{MB} ~ .
\label{DLphoto}
\eeq
I will not discuss the chiral expansion of $E_{0+}$ step by step, referring
to the literature \cite{BGKM} \cite{BKM1} \cite{BKKM} for the actual
calculation and for more details to the Comment
\cite{gerulf}. Up--to--and--including order $\mu^2$, one has to
consider contributions with $D_L = 0, 1$ and $2$. In fact, for neutral
pion photoproduction, there is no term with $D_L=0$ since the
time--honored Kroll--Ruderman contact term
\cite{KR} where both the pion and the photon emanate from the same
vertex, only exists for charged pions.
For $D_L = 2$
there is a one--loop
contribution ($L=1$) with leading--order vertices only
($N_d^{MB}=0 ~(d>1)$).  It is considerably easier to work out the
relevant diagrams in HBCHPT \cite{BKKM} than in the original derivation
\cite{BGKM} \cite{BKM1}. In fact, at threshold only the so--called triangle
diagram (and its crossed partner) survive out
of some 60 diagrams. The main reason for the enormous simplification
in HBCHPT is that one can choose a gauge without a direct $\gamma NN$
coupling of lowest order and that there is no direct coupling of
the produced $\pi^0$ to the nucleon at threshold.
Notice that the loop contributions are finite and they are identical for proton
and neutron. They were omitted in the original version of the LET
\cite{VZ} \cite{deB} and in many later rederivations.
The full LETs of ${\cal O}(p^3)$ are given by \cite{BGKM}
\beq
E_{0+}(\pi^0 p)=  - \dfrac{e g_A}{8\pi F_\pi}\biggl[ \biggr.
\dfrac{M_\pi}{m} - \dfrac{M_\pi^2}{2 m^2}~(3+\kappa_p)  -
\dfrac{M_\pi^2}{16 F_\pi^2} + \biggl.{\cal O}(M_\pi^3)\biggr] \eeq
\beq
E_{0+}(\pi^0 n)=  - \dfrac{e g_A}{8\pi F_\pi}\biggl[ \biggr.
 \dfrac{M_\pi^2}{2 m^2}~\kappa_n  -
\dfrac{M_\pi^2}{16 F_\pi^2} + \biggl. {\cal O}(M_\pi^3)\biggr] \eeq
We note that the electric dipole amplitude for neutral pion production
vanishes in the chiral limit. By now, even the terms of order
$M_\pi^3$ have been worked out, see ref.\cite{bkmpi0}.

The derivation of LETs sketched above is based on a well--defined
quantum field theory where each step can be checked explicitly. Nevertheless,
the corrected LETs have been questioned by several authors. We
find it instructive to discuss some of the arguments and assumptions
that have been used to derive or rederive the original LETs.
(a) {\it Analyticity} {\it  assumption:}
The original derivations \cite{VZ} \cite{deB} and some later rederivations
\cite{Naus1} \cite{Naus2} \cite{Kamal} \cite{bh} used a Taylor expansion
of amplitudes in the variables $\nu, \nu_B$ (linear combinations of the
usual Mandelstam variables $s$ and $u$). The seemingly plausible assumption
that the coefficients of this expansion are analytic in $M_\pi$ leads
directly to the original low--energy guess (LEG).
In fact, in Ref.~\cite{VZ} it was explicitly
spelled out that this is a necessary assumption for the LEG
to hold. However, as shown in
Ref.~\cite{BGKM}, this assumption does not hold in QCD. Due to the Goldstone
nature of the pion, some Taylor coefficients diverge in the chiral
limit. This happens precisely in the loop contributions ($D_L=2$)
which generate infrared divergences in some coefficients.
The threshold amplitude itself is perfectly well--behaved in the chiral limit.
(b) {\it External} {\it versus} {\it  internal} {\it  pion} {\it  mass:}
It has been suggested \cite{Naus2} \cite{Naus3} \cite{Scherer}
that there is a basic difference between the external,
kinematical pion mass $M_\pi$ and the internal mass $\bar M_\pi$
appearing in the pion propagators in loop diagrams. The assumption
is that $\bar M_\pi$ appears only in relations between unrenormalized
and renormalized quantities. Therefore, expressing everything in
measurable, renormalized quantities, no trace of $\bar M_\pi$ is
left and one recovers the original LEG since the loop
contribution is to be dropped by assumption.
In HBCHPT the difference between the chiral limit values $F,
\krig{g}_A , M , \krig{\kappa}_p$ and the physical parameters
$F_\pi, g_A , M_\pi , \kappa_p$
only affects terms of ${\cal O}(M_\pi^3)$ in
the LETs for $\pi^0$ photoproduction at threshold. Thus, the loop
contribution to the LETs cannot be a renormalization effect.
There is a more fundamental objection to the distinction between external
and internal pion masses. QCD does not offer a consistent
procedure for $M_\pi \to 0$ with $\bar M_\pi$ remaining finite. The
only tunable mass parameters in QCD are the quark masses.
Letting the quark masses tend to zero makes all pion masses vanish,
whether they be external or internal.
(c){\it  Off--shell} {\it  expansion:}
The inadmissible distinction between external and internal pion masses
can also appear in an off--shell extrapolation of the amplitude.
Davidson has contrasted the expansion in $M_\pi$ with
a so--called $\omega$ expansion \cite{David}. Keeping $M_\pi$ fixed, he
sets the three--momentum $\vec p_\pi =0$ and expands in the pion
energy $E_\pi = \omega$. Obviously, for $\omega \ne M_\pi$ this
implies an off--shell extrapolation of the scattering amplitude.
If one expands the amplitude first to ${\cal O}(\omega^2)$, the coefficients
still depend on $M_\pi$. Expanding those coefficients in a second step
in $M_\pi$ so that the overall order is ${\cal O}(M_\pi^2)$ for
$\omega = M_\pi$, one obtains the original LEG \cite{David}. The
mathematical origin of the problem is an illicit interchange of limits~:
expanding a function $f(\omega,M_\pi)$ in the manner just described
and setting $\omega = M_\pi$ at the end will in general not lead
to the same result as an expansion of $f(M_\pi,M_\pi)$ to the
same order in $M_\pi$.
Although it is shown in Ref.~\cite{David} that one can recover the
correct LET by a resummation of the series to all orders in $\omega$,
there is in general no guarantee that off--shell manipulations
produce the correct result. A simple, but
instructive example is to consider the elastic $\pi\pi$
scattering amplitude to lowest order, ${\cal O}(p^2)$, both in CHPT and
in the linear $\sigma$ model. Although the amplitudes agree on--shell,
they disagree in general off--shell. In fact, one can obtain very
different forms for the off--shell amplitude by redefining
the pion field. While one would normally not employ such
redefinitions in the linear model (seemingly destroying
renormalizability), any choice of pion field is equally acceptable
in CHPT which is based on an intrinsically non--renormalizable
quantum field theory.
Off--shell manipulations are dangerous and may lead to incorrect
results. The literature on applications of current algebra techniques
abounds with examples. Concerning the
{\it phenomenology} of neutral pion photoproduction off nucleons, I
refer the reader to the extensive discussion in
ref.\cite{bkmpi0}. There, it is shown that the electric dipole
amplitude $E_{0+}$ is indeed not a good testing ground of the chiral
dynamics but rather there exist LETs in the {\it P--waves}, which can be
used to help analyze the data. An example is given in the next
section.

\bigskip

The last example I want to treat is
the case of two--pion photoproduction. At threshold, the
two--pion photoproduction current matrix element can be decomposed
into amplitudes as follows (to first order in $e$ and in the gauge $
\epsilon_0=0$):
\beq T \cdot \epsilon = \chi^\dagger_f \left\lbrace i \vec \sigma \cdot
(\vec \epsilon \times \vec k \,) [ M_1 \delta^{ab} +
M_2 \delta^{ab} \tau^3 + M_3 (\delta^{a3} \tau^b + \delta^{b3} \tau^a)]
\right\rbrace \, , \eeq
with $\chi_{i,f}$ two--component Pauli and isospinors. Here, I am only
interested in the first non--vanishing contributions to $M_{2,3}$,
given by some tree diagrams. So we have $N_\gamma =1$ , $L=0$ and only
lowest order mesonic vertices ($d=2$), i.e.
\beq D_L = \sum_d (d-1) N_d^{MB} \quad . \eeq
Tree diagrams with lowest order vertices from ${\cal L}_{\pi N}^{(1)}$
all vanish due to the threshold selection rules $\epsilon \cdot v =
\epsilon \cdot q_i =0$, $S \cdot q_i = 0$ and $v \cdot (q_1 -q_2) =
0$. The first non--vanishing contribution comes from tree diagrams
with insertions from ${\cal L}_{\pi N}^{(2)}$, in particular the
expansion of $f_{\mu \nu}^+$ from eq.(\ref{lp2}) leads to a $\gamma
\pi \pi NN$ vertex proportional to $\krig{\kappa}_V$ so that
\beq M_2 = -2M_3 = \frac{e}{4 \krig{m} F^2} \left( 2 \krig{g}_A^2 - 1 -
\krig{\kappa}_V \right) + {\cal O}(p)
= \frac{e}{4 \ m F_\pi^2} \left( 2 g_A^2 - 1 -
\kappa_V \right) + {\cal O}(p) \, \, . \eeq
The full one--loop results can be found in ref.\cite{bkm2p}. It is
amusing to note that this particular vertex has been overlooked in
other  derivations of these LETs \cite{dd} \cite{bt} simply because
not the most {\it general} effective $\pi N$ Lagrangian at order $p^2$
was used.

\section{Confronting the data}
\label{sec:data}

As already stated in the introduction, most of the precision data on
the nucleon at low energies come from processes involving real or
virtual photons such as pion photo-- and electroproduction as well as
Compton scattering. This is mostly due to the advent of the CW
accelerators such as MAMI at Mainz and improved detector
technology. It is worth to stress that there are on--going activities
in these fields also at NIKHEF (Amsterdam), SAL (Saskatoon), BATES
(MIT), ELSA (Bonn) and other places. Other precise data come from
atomic energy shifts at PSI (Villigen)   and there is also a
tremendous amount of threshold data for $\pi N \to \pi \pi N$ from such
places like TRIUMF (Vancouver). Just from the beginning I would like to
stress that one should not see these different experiments and data in
isolation but that they are rather intimately connected. For example,
the imaginary parts of the various multipoles in pion photo-- and
electroproduction are proportional to the respective pion--nucleon
scattering phase shifts via the Fermi--Watson final state theorem. As
an example, let me give the imaginary part of the electric dipole
amplitude in neutral pion photoproduction off protons,
\beq {\rm Im}~E_{0+}^{\pi^0 p} = {\rm Re}~E_{0+}^{(0)} \tan(\delta_{1/2}) +
\frac{1}{3}{\rm Re}~E_{0+}^{(1/2)} \tan(\delta_{1/2}) +
\frac{2}{3}{\rm Re}~E_{0+}^{(3/2)} \tan(\delta_{3/2})  \, \, , \eeq
with $\delta_{1/2,3/2}$ the pion--nucleon S--wave phase shifts for
total isospin 1/2 and 3/2, respectively. Also, various contact terms
show up in different reactions. For example, the LECs $c_{1,2,3}$
which can be determined in $\pi N \to \pi N$ \cite{bkmpin} show up in
the order $q^4$ calculation of the nucleons' electromagnetic
polarizabilities \cite{bkms} from non--vanishing insertions of ${\cal
  L}_{\pi N}^{(2)}$. This should always be kept in mind. For a nice
and instructive flow--chart giving the links between low--energy pion
experiments I refer the reader to Fig.7.33--1 in the monograph of
de~Benedetti \cite{debe}.

Let us now consider some of the data with electromagnetic and weak
probes which have become available over the last few years. In
table~2, I summarize these in comparison to the chiral predictions.
Clearly, there are more measured quantities, but I have only included
the ones which (a) can be predicted with some accuracy and (b) for
which some solid data exist. The most notabel absentee from this table
is the electric dipole amplitude for $\gamma p \to \pi^0 p$. First, as
discussed in length in refs.\cite{vero} \cite{bkmpi0},
no solid chiral prediction
exists at present and second, the empirical determination hinges on a
few empirical points and some more or less justified assumptions about
the P--waves. This reaction has been remeasured at MAMI and SAL and
one should wait for the analysis of these new data. Quite in contrast
to common folklore, there exist a set of LETs for the slopes of the P--waves
$P_{1,2} = 3E_{1+} \pm M_{1+} \mp M_{1-}$ at threshold, e.g.
\beq \frac{1}{|\vec q \,|} P_{1, {\rm thr}}^{\pi^0 p} = \frac{e g_{\pi
  N}}{8 \pi m^2} \left\lbrace 1 + \kappa_p + \mu \left[ -1 -
\frac{\kappa_p}{2} + \frac{g_{\pi N}^2}{48 \pi}(10 -3\pi) \right]
+ {\cal O}(\mu^2) \right\rbrace \, \, . \eeq
Numerically, this translates into
\beq \frac{1}{|\vec q \,|} P_{1, {\rm thr}}^{\pi^0 p} = 0.512 \, ( 1
-0.062) \, {\rm GeV}^{-2} = 0.480 \, {\rm GeV}^{-2} \, \, , \eeq
which is given in table~2 in units which are more used in the literature.
For a discussion of the other quantities, I refer to the quoted references.

\begin{center}

\begin{tabular}{|ccccccc|} \hline
Observable & Prediction &  Order  & Ref. & Data & Ref. & Units  \\ \hline
 & & & & & & \\
$\bar{\alpha}_p$ & $10.5\pm 2.0$  & $q^4$& \cite{bkms} & $10.4\pm0.6$
& \cite{feder} \cite{sal} & 10$^{-4}$ fm$^{3}$  \\
$\bar{\beta}_p$ & $3.5\pm 3.6$  & $q^4$& \cite{bkms} & $3.8\mp0.6$
& \cite{feder} \cite{sal} & 10$^{-4}$ fm$^{3}$  \\
$\bar{\alpha}_n$ & $13.4\pm 1.5$  & $q^4$& \cite{bkms} & $12.3\pm1.3$
& \cite{schmi} & 10$^{-4}$ fm$^{3}$  \\
$\bar{\beta}_n$ & $7.8\pm 3.6$  & $q^4$& \cite{bkms} & $3.5\mp 1.3$
& \cite{schmi} & 10$^{-4}$ fm$^{3}$  \\
$E_{0+,{\rm thr}}^{\pi^+ n}$ & $ 28.4$  & $q^3$& \cite{BKM1} &
$27.9\pm 0.5$  & \cite{burg} & 10$^{-3}/ M_{\pi^+}$ \\
$E_{0+,{\rm thr}}^{\pi^- p}$ & $ -31.1$  & $q^3$& \cite{BKM1} &
$-31.4\pm 1.3$  & \cite{burg} & 10$^{-3}/ M_{\pi^+}$ \\
$P_{1,{\rm thr}}^{\pi^0 p} $ & $ 10.9$  & $q^3$& \cite{bkmpi0} &
$8.8\pm 0.6$  & \cite{dt} & $|\vec q \, ||\vec k \,|$ 10$^{-3}/M_{\pi^+}^3$ \\
$|L_{0+,{\rm thr}}^{\pi^0 p}|^2$ & $ 0.20$  & $q^3$& \cite{bklm} &
$0.13\pm 0.05$  & \cite{pat} & $\mu$b \\
$ g_P $ & $ 8.44 \pm 0.23$  & $q^3$& \cite{bkmgp} &
$8.7\pm 1.6$  & \cite{richter} & --  \\ & & & & & &
 \\ \hline \end{tabular}
\smallskip

\end{center}

\noindent Table 2: Chiral predictions for electroweak probes
compared to experimental data. Given are the electric ($\bar \alpha$) and
magnetic ($\bar \beta$) polarizabilities of the proton and the neutron, the
electric dipole amplitudes for charged pion photoproduction, the P--wave
$P_1$ for neutral pion photoproduction off protons, the longitudinal
S--wave multipole for neutral pion electroproduction off protons and
the induced pseudoscalar coupling constant, $g_P$, measured in muon capture.
In some cases, the
theoretical uncertainty of the calculation has been estimated and the
respective order ($q^3$ or $q^4$) is also given.

 \bigskip

There is, of course, ample of data on elastic pion--nucleon scattering
and on $\pi N \to \pi \pi N$ in the threshold region. For a review on
the status in $\pi N \to \pi N$, I refer to Sainio
and H\"ohler \cite{mikko} \cite{hoehmit}. Here,
I briefly want to mention some work concerning the chiral corrections
to these processes in the framework of HBCHPT. Consider first the
on--shell forward $\pi N$ scattering amplitude \cite{bkmpin},
\beq T = \delta^{ab} \, T^+ (\omega) + i \, \epsilon^{bac} \, \tau^c
\, T^- (\omega) \, \, , \eeq
with $\omega = v \cdot q$ the pion energy. At threshold, $\omega =
M_\pi$, we have
\beq T^- (M_\pi) = 4 \pi (1 + \mu) \, a^- = \frac{4}{3} \pi (1 + \mu)
\, (a_{1/2} - a_{3/2}) \, \, , \eeq
with $\mu = M_\pi / m$ and $a_{1/2, 3/2}$ are the S--wave scattering
lengths in the channels with total isospin 1/2 and 3/2, respectively.
The chiral expansion of $T^- (M_\pi)$ reads:
\beq T^- (M_\pi) = \frac{M_\pi}{2 F_\pi^2} + \frac{g_{\pi N}^2
  M_\pi^3}{8 m^4} + \frac{M_\pi^3}{16 \pi^2 F_\pi^4} \left( 1 - 2 \ln
 \frac{M_\pi}{\lambda} \right)+ \frac{g_{\pi N}^2  M_\pi^3}{2 m^2
   m^2_\Delta} (Z- \frac{1}{2})^2 + {\cal O}(M_\pi^5) \, \, . \label{tpi} \eeq
It is remarkable that there are no corrections of order $M_\pi^2$ and
$M_\pi^4$. The various terms in eq.(\ref{tpi}) are the current algebra
prediction \cite{wein66}, the expansion of the nucleon pole term, the one--loop
and the counterterm contribution form ${\cal L}_{\pi N}^{(3)}$,
respectively. $\lambda$ is the scale introduced in dimensional
regularization and the pertinent LEC has been estimated by $\Delta$
exchange (there is also a small contribution from the Roper which I drop).
At the resonance scale $\lambda = m_\Delta$, we find
\beq T^- (M_\pi) = (1.59 + 0.02 + 0.24 + 0.04(Z - 1/2)^2) \, {\rm fm}
= 1.87 \, {\rm fm}              \eeq
for the $\Delta N \pi$ off-shell parameter $Z = -1/4$. This agrees
nicely with the empirical value of $T^- (M_\pi) = (1.86 \pm 0.04)$ fm
\cite{land}.
If one chooses $\lambda = m$, the contribution from the loops drops to
0.22 fm. The main message here is that {\it only} the {\it loop}
{\it  correction} at
order $M_\pi^3$ can close the gap between the Weinberg--Tomozawa
prediction \cite{wein66} \cite{tomo} and the empirical
value. Furthermore, the contact term contribution is small so
uncertainties in the corresponding LECs get washed out by a small prefactor.

\bigskip

The last issue I want to address is the reaction $\pi N \to \pi \pi
N$. Here, ample new and rather accurate data
in the threshold region exist
\cite{poc} \cite{bl} \cite{sev} \cite{kern} \cite{lowe}
 but the one--loop calculation has not yet been
performed. However, in ref.\cite{bkmppn} the first corrections to the
lowest order where considered. At threshold, the T--matrix can be
written in terms of two  amplitudes $D_1$ and $D_2$,
\beq T = i \vec \sigma \cdot \vec k \left[ D_1 (\tau^b \delta^{ac}
+ \tau^c \delta^{ab}) \, + \, D_2 \tau^a \delta^{bc} \right] \, \, , \eeq
where $a,b,c$ are isospin indices of the pions and $\vec k$ denotes
the three momentum of the incoming pion. One can then derive LETs for
$D_{1,2}$ \cite{bkmppn}
\beq D_1 = \frac{g_A}{8 F_\pi^3} \left( 1 +\frac{7 M_\pi}{2m} \right)
+ {\cal O}(M_\pi^2)  = 2.4 \, \, {\rm fm}^3 \,\, , \label{lppn1}\eeq
\beq D_2 = -\frac{g_A}{8 F_\pi^3} \left( 3 +\frac{17 M_\pi}{2m} \right)
+ {\cal O}(M_\pi^2)  = -6.8 \, \, {\rm fm}^3 \,\,  \label{lppn2} \eeq
Now the amplitude $D_1$ is exclusively sensitive to the two--pion
final state with total isospin $I_{\pi \pi} =2$ whereas $D_2$ is dominated
by the isospin zero S--wave. Consequently, one expects  a small and a
sizeable correction at order $M_\pi^2$ for $D_1$ and $D_2$, in
order. This is corroborated by the calculation of the imaginary part of
the one--loop diagrams which give a non--zero imaginary part at that
order as detailed in ref.\cite{bkmppn}. A best fit to the  data
from threshold to 20 MeV above (for the pion kinetic energy) gives the
empirical values
\beq D_1 = 2.2 \, {\rm fm}^3 \, , \quad D_2 = -8.8 \, {\rm fm}^3 \, , \eeq
which are $9\%$ and $29\%$ off the LET predictions,
eqs.(\ref{lppn1},\ref{lppn2}).
These deviation show exactly the trend discussed before. Clearly, it
is mandatory to complete the full order $q^3$ calculation before any
further conclusions can be drawn.

\section{Problems and open questions}

Here, I will list a few topics which deserve further study. This list
should neither be considered complete nor does the ordering imply any
priority.

\begin{enumerate}
\item[(i)] In the two--flavor sector, we have to perform the one--loop
calculation for $\pi N \to \pi \pi N$ (to order $q^3$ or even better, $q^4$).
At present, only tree level calculations and the first corrections from
${\cal L}_{\pi N}^{(2)}$ are available. This is in marked contrast  to
the many existing rather precise data. What one finally wants to learn
is to  what {\it precision} these threshold pion production data encode
information about the low energy elastic $\pi \pi$ scattering
amplitude. Presently available determimations of the S--wave $\pi \pi$
scattering lenghts from these data should only be considered as estimates.

\smallskip

\item[(ii)] More precise data to which HBCHPT can be applied are
needed. This would then allow for a systematic study of the LECs
appearing in ${\cal L}_{\pi N}^{(3,4)}$ and to judge the valitidy
(quality) of the resonance saturation principle which is often used to
get a handle on the LECs. This calls for a joint effort of the
experimenters and the theoreticians.

\smallskip

\item[(iii)] In the three flavor sector, despite lots of mumbling and talking,
we do not yet have a {\it consistent} picture or a set of complete (and
hopefully accurate) calculations to draw decisive conclusions. There
exist many data and more are coming, as an example let me just mention the
accurate threshold kaon photoproduction ones from ELSA or the
proposals to measure the hyperon polarizabilities at Fermilab and
CERN.

\smallskip

\item[(iv)] Jenkins and Manohar  first advocated to supplement the EFT
of the ground state baryons and Goldstone bosons by the spin-3/2
decuplet \cite{jmdel} \cite{dobo}. This approach has been taken up in
quite a few papers thereafter. If one thinks of extending calculations
like $\pi N \to \pi N$ through the $\Delta$ region, this is certainly
unavoidable. However, what is missing is a set of {\it complete}
calculations for threshold observables from which one could (a) assess
the accuracy of the approach and (b) the residual octet--decuplet mass
difference has to be treated in a more systematic fashion.

\smallskip

\item[(v)] The extension of the CHPT approach to systems  of two or
more nucleons has only begun. Despite some theoretical problems (the
power counting only applies to the subset of irreducible diagrams),
it seems to shed some light on the phenomenology of nuclear forces like
e.g. the smallness of three--, four-- , $\dots$ body forces and the
masking of isospin violation in these systems. However, it is
mandatory to perform these calculations with {\it all} the input which
is available from the single baryon sector. This has not yet been
done.

\end{enumerate}

\bigskip

\section{Acknowledgements}
First, I would like to thank the organizers for their kind invitation
and hospitality. I am also grateful to  V\'{e}ronique
Bernard and Norbert Kaiser for fruitful collaborations and allowing me
to present some material before publication. I would like to thank
Gerhard Ecker and J\"urg Gasser for sharing with me their insight into
the chiral dynamics.


\bigskip

\baselineskip 14pt

\begin{thebibliography}{99}

\bibitem{sac} E. Mazzucato et al., Phys. Rev. Lett. {\bf 57} (1986) 3144

\bibitem{beck} R. Beck et al., Phys. Rev. Lett. {\bf 65} (1990) 1841

\bibitem{pat} T.P. Welch et al., Phys. Rev. Lett. {\bf 69} (1992) 2761

\bibitem{feder} F. Federspiel et al., Phys. Rev. Lett. {\bf 67} (1991) 1511

\bibitem{schmi} J. Schmiedmayer et al., Phys. Rev. Lett. {\bf 66} (1991) 1015

\bibitem{sal} E.L. Hallin et al., Phys. Rev.  {\bf C48} (1993) 1497

\bibitem{zieger} A. Zieger et al., Phys. Lett. {\bf B278} (1992) 34

\bibitem{alfons} A. Buchmann, these proceedings

\bibitem{GL1} J. Gasser and H. Leutwyler, Ann. Phys. {\bf 158} (1984) 142

\bibitem{GL2} J. Gasser and H. Leutwyler, Nucl. Phys. {\bf B250} (1985) 465

\bibitem{UGM} Ulf-G. Mei{\ss}ner, Rep. Prog. Phys. {\bf 56} (1993) 903

\bibitem{ecker} G. Ecker, Czech. J. Phys. {\bf 44} (1994) 405

\bibitem{col}
S. Coleman, ``Aspects of Symmetry", Cambridge University Press,
Cambridge, 1985

\bibitem{wein79} S. Weinberg, Physica {\bf 96A} (1979) 327

\bibitem{GSS} J. Gasser, M.E. Sainio and A. $\rm{\check S}$varc,
Nucl. Phys. {\bf B307} (1988) 779

\bibitem{JM} E. Jenkins and A.V. Manohar, Phys. Lett. {\bf B255} (1991) 558

\bibitem{leut} H. Leutwyler, Ann. Phys. {\bf 235} (1994) 165

\bibitem{ccwz} S. Coleman, J. Wess and B. Zumino, Phys. Rev. {\bf 177}
(1969) 2239; C.G. Callan, S. Coleman, J. Wess and B. Zumino, {\it
ibid} 2247

\bibitem{weinno} S. Weinberg, Phys. Rev. {\bf 177} (1968) 1568

\bibitem{BKKM}
V. Bernard, N. Kaiser, J. Kambor and Ulf-G. Mei\ss ner, Nucl. Phys.
{\bf B388} (1992) 315

\bibitem{GTR} M. Goldberger and S.B. Treiman, Phys. Rev. {\bf 110} (1958) 1478

\bibitem{rho} M. Rho, Phys. Rev. Lett. {\bf 66} (1991) 1275

\bibitem{weinnn} S. Weinberg, Phys. Lett. {\bf B251} (1990) 288

\bibitem{weinnp} S. Weinberg, Nucl. Phys. {\bf B363} (1991) 3

\bibitem{bkms}
V. Bernard, N. Kaiser, Ulf-G. Mei\ss ner and A. Schmidt, Phys. Lett.
{\bf B319} (1993) 269; Z. Phys. {\bf A348} (1994) 317

\bibitem{dobo} E. Jenkins and A.V. Manohar, in ``Effective Field
Theories of the Standard Model'', ed. Ulf-G. Mei{\ss}ner, World
Scientific, Singapore, 1992

\bibitem{coll} J.C. Collins, ``Renormalization'', Cambridge University
Press, Cambridge, 1984

\bibitem{lapa} P. Langacker and H. Pagels, Phys. Rev. {\bf D8} (1975) 4595


\bibitem{Stern}
N.H. Fuchs, H. Sazdjian and J. Stern, Phys. Lett. {\bf B269} (1991) 183;\\
J. Stern, H. Sazdjian and N.H. Fuchs, Phys. Rev. {\bf 47} (1993) 3814

\bibitem{gass} J. Gasser, Ann. Phys. {\bf 136} (1981) 62

\bibitem{liz} E. Jenkins, Nucl. Phys. {\bf B368} (1992) 190

\bibitem{jms} E. Jenkins and A.V. Manohar, Phys. Lett. {\bf B281} (1992) 336

\bibitem{bkmz}
V. Bernard, N. Kaiser and Ulf-G. Mei\ss ner, Z. Phys. {\bf C60} (1993) 111

\bibitem{ll} R.F. Lebed and M.A. Luty, Phys. Lett. {\bf B29} (1994) 479

\bibitem{samir} S. Mallik, ``Massive States in Chiral Perturbation
 Theory'', Saha Institute preprint, 1994

\bibitem{mary} M.K. Banerjee and J. Milana,
  ``Baryon Mass Splittings in Chiral Perturbation
 Theory'', preprint UMPP 95-058, 1994

\bibitem{bkmpin}
V. Bernard, N. Kaiser and Ulf-G. Mei\ss ner, Phys. Lett. {\bf B309} (1993) 421

\bibitem{bkmrev} V. Bernard, N. Kaiser and Ulf-G. Mei\ss ner,
``Chiral Symmetry in Nuclear Physics'', in preparation

\bibitem{bira} C. Ordonez and U. van Kolck,  Phys. Lett. {\bf B291} (1992) 459;
C. Ordonez, L. Ray and U. van Kolck,  Phys. Rev. Lett. {\bf 72} (1994)
1982;  U. van Kolck,  Phys. Rev. {\bf C49} (1994) 2932

\bibitem{birat} U. van Kolck, thesis, Univ. of Texas at Austin, 1993

\bibitem{reso} G. Ecker, J. Gasser, A. Pich and E. de Rafael,
Nucl. Phys. {\bf 321} (1989) 311; J.F. Donoghue, C. Ramirez and
G. Valencia, Phys. Rev. {\bf D39} (1989) 1947

\bibitem{ulfmit}
Ulf-G. Mei{\ss}ner, ``Aspects of Nucleon Chiral
Perturbation Theory", talk given at the Workshop on Chiral Dynamics~:
Theory and Experiment, MIT, Cambridge, July 1994, preprint CRN 94/44

\bibitem{eckp3} G. Ecker. Phys. Lett. {\bf B336} (1994) 508

\bibitem{bkmpi0}V. Bernard, N. Kaiser and Ulf-G. Mei\ss ner, ``Neutral
Pion Photoproduction off Nucleons Revisited'', preprint CRN 94-62
and TK 94 18, 1994

\bibitem{gerulf}G. Ecker and  Ulf-G. Mei\ss ner, ``What is a
  Low--Energy Theorem?'', prperint CRN 94--52 and UWThPh-1994-33, 1994

\bibitem{low} F. Low, Phys. Rev. {\bf 96} (1954) 1428;\\
M. Gell-Mann and M.L. Goldberger, ibid. 1433

\bibitem{wein68} S. Weinberg, Phys. Rev. Lett. {\bf 18} (1967) 188

\bibitem{BKM}
V. Bernard, N. Kaiser and Ulf-G. Mei\ss ner, Nucl. Phys. {\bf B373} (1992) 346

\bibitem{BGKM}
V. Bernard, J. Gasser, N. Kaiser and Ulf-G. Mei\ss ner, Phys.
Lett. {\bf B268} (1991) 291

\bibitem{BKM1}
V. Bernard, N. Kaiser and Ulf-G. Mei\ss ner, Nucl. Phys. {\bf B383} (1992) 442

\bibitem{KR}
N.M. Kroll and M.A. Ruderman, Phys. Rev. {\bf 93} (1954) 233

\bibitem{VZ} A.I. Vainsthein and V.I. Zakharov, Yad. Fiz. {\bf 12}
(1970) 610 (Sov. J. Nucl. Phys. {\bf 12} (1971) 333); Uspekhi
Fiz. Nauk {\bf 100} (1970) 225 (Sov. Phys. Uspekhi {\bf 13}
(1970) 73);  Nucl. Phys. {\bf B36} (1972) 589

\bibitem{deB} P. de Baenst, Nucl. Phys. {\bf B24} (1970) 633

\bibitem{bh} A.M. Bernstein and B.R. Holstein,
Comments Nucl. Part. Phys. {\bf 20} (1991) 197

\bibitem{Naus1}
H.W.L. Naus, J.H. Koch and J.L. Friar, Phys. Rev. {\bf C41} (1990) 2852

\bibitem{Naus2}
H.W.L. Naus, Phys. Rev. {\bf C43} (1991) R365

\bibitem{Kamal}
A.N. Kamal, Int. J. Mod. Phys. {\bf A6} (1991) 263

\bibitem{Naus3}
H.W.L. Naus, Phys. Rev. {\bf C44} (1991) 531

\bibitem{Scherer}
S. Scherer, J.H. Koch and J.L. Friar, Nucl. Phys. {\bf A552} (1993) 515

\bibitem{David}
R.M. Davidson, Phys. Rev. {\bf C47} (1993) 2492

\bibitem{bkm2p}
V. Bernard, N. Kaiser and Ulf-G. Mei\ss ner, Nucl. Phys. {\bf A580}
(1994) 475

\bibitem{dd} R. Dahm and D. Drechsel, in Proc. Seventh Amsterdam
Mini-Conference, eds. H.P. Blok, J.H. Koch and H. De Vries, Amsterdam,
1991

\bibitem{bt} M. Benmerrouche and R. Tomusiak, Phys. Rev. Lett. {\bf
  73} (1994) 400

\bibitem{debe} S. DeBenedetti, ``Nuclear Interactions'', John Wiley
  and Sons, New York, 1964

\bibitem{vero}
V. Bernard, ``Threshold Pion Photo-- and Electroproduction in Chiral
Perturbation Theory", talk given at the Workshop on Chiral Dynamics~:
Theory and
Experiment, MIT, Cambridge, July 1994, preprint CRN 94/45, hep-ph/9408323

\bibitem{burg} J. P. Burg, Ann. Phys. (Paris) {\bf 10} (1965) 363

\bibitem{dt} D. Drechsel and L. Tiator, J. Phys. G:
  Nucl. Part. Phys. {\bf 18} (1992) 449

\bibitem{bklm} V. Bernard, T.--S. H. Lee, N. Kaiser
and Ulf-G. Mei{\ss}ner,  Phys. Rep. {\bf 246} (1994) 315

\bibitem{bkmgp}V. Bernard, N. Kaiser and Ulf-G. Mei\ss ner,
  Phys. Rev. {\bf D50} (1994) in print

\bibitem{richter}G. Bardin et al., Phys. Lett. {\bf B104} (1981) 320

\bibitem{land} G. H\"ohler, Landolt--B\"ornstein vol. 9b2,
H. Schopper (ed.), Springer,  Berlin, 1983

\bibitem{mikko} M. E. Sainio, ``Pion--Nucleon Sigma Term'',
talk given at the Workshop on Chiral Dynamics~: Theory and
Experiment, MIT, Cambridge, July 1994, preprint HU-TFT-94-40

\bibitem{hoehmit} G. H\"ohler, ``Tests of Predictions from Chiral
Perturbation Theory for $\pi N$ Scattering'',
talk given at the Workshop on Chiral Dynamics~: Theory and
Experiment, MIT, Cambridge, July 1994, Karlsruhe preprint


\bibitem{wein66} S. Weinberg, Phys. Rev. Lett. {\bf 17} (1966) 616

\bibitem{tomo} Y. Tomozawa, Nuovo Cim. {\bf 46A} (1966) 707

\bibitem{poc} D. Pocancic et al., Phys. Rev. Lett. {\bf 72} (1994) 1156

\bibitem{bl} H. Burkhardt and J. Lowe, Phys. Rev. Lett. {\bf 67} (1991) 2622

\bibitem{sev} M.E. Sevior et al., Phys. Rev. Lett. {\bf 66} (1991) 2569

\bibitem{kern} G. Kernel et al., Z. Phys. {\bf C48} (1990) 201

\bibitem{lowe} J. Lowe et al., Phys. Rev.  {\bf C44} (1991) 956

\bibitem{bkmppn}
V. Bernard, N. Kaiser and Ulf-G. Mei\ss ner, Phys. Lett. {\bf B332}
(1994) 415

\bibitem{jmdel} E. Jenkins and A.V. Manohar, Phys. Lett. {\bf B259} (1991) 353


\end{thebibliography}
\end{document}